\documentclass[aps,preprint]{revtex4}
\usepackage{amsmath}
\usepackage{amssymb}
\usepackage[italian,english]{babel}
\usepackage{subfigure}
\usepackage{mathrsfs}
\usepackage[dvips]{graphicx}

\renewcommand{\deg}{\rm ^o}
\newcommand{\rAA}{\rm \AA}

\begin{document}
\title{Use of the Metropolis algorithm to simulate the dynamics of protein chains}
\author{G. Tiana$^{1,2}$, L. Sutto$^{1,2}$ and R. A. Broglia$^{1,2,3}$}
\affiliation{$^{1}$Dipartimento di Fisica, Universit\'a di Milano, via Celoria 16, 20133 Milano, Italy}
\affiliation{$^{2}$INFN, Sez. di Milano, Milano, Italy}
\affiliation{$^{3}$Niels Bohr Institute, University of Copenhagen, Bledgamsvej 17, 2100 Copenhagen, Denmark}
\date{\today}
\begin{abstract}
The Metropolis implementation of the Monte Carlo algorithm has been developed to study the equilibrium thermodynamics of many--body systems. Choosing small trial moves, the trajectories obtained applying this algorithm agree with those obtained by Langevin's dynamics. Applying this procedure to a simplified protein model, it is possible to show that setting a threshold of 1$\deg$ on the movement of the dihedrals of the protein backbone in a single Monte Carlo step, the mean quantities associated with the off--equilibrium dynamics (e.g., energy, RMSD, etc.) are well reproduced, while the good description of higher moments requires smaller moves. An important result is that the time duration of a Monte Carlo step depends linearly on the temperature, something which should be accounted for when doing simulations at different temperatures.
\end{abstract}
\maketitle

\section{Introduction}

Since its publication by Metropolis, Rosenbluth, Teller and their wives in 1953, the algorithm designed for ``equation of state calculations by fast computing machines" \cite{metropolis} has been used to obtain an approximation of the equilibrium properties of a wide range of classical systems. Of course, in fifty years, fast computing machines has become faster and faster, and this decreed the success of the algorithm. 

The Metropolis algorithm performs a sample of the configuration space of a system starting from a random conformation and repeating a large number of steps. Each step consists of attempting a transition to a new conformation $x'$ choosing among a set of allowed moves, and accepting the attempt with probability $\min[1,\exp(-(U(x)-U(x'))/T)]$ where $U$ is the potential energy and T the absolute temperature in units of Boltzmann's constant. This is equivalent to solve numerically the master equation
\begin{equation}
\frac{\partial p(x,t)}{\partial t}=\int dx'\; \left[p(x',t)w(x'\rightarrow x)-p(x,t)w(x\rightarrow x')\right].
\label{masterequation}
\end{equation}
where the transition rates are
\begin{equation}
w(x'\rightarrow x)=w_0\cdot p_{ap}(x'\rightarrow x)\cdot \min\left[1,\exp\left(-\frac{U(x)-U(x')}{T}\right)\right],
\label{metropolis}
\end{equation}
where $w_0$ sets the time scale of the transitions and $p_{ap}(x'\rightarrow x)$ is the {\it a priori} probability of choosing the move which goes from the state $x'$ to the state $x$. If the {\it a priori} probability satisfies $p_{ap}(x'\rightarrow x)=p_{ap}(x\rightarrow x')$ and allows the system to visit the whole phase space, the algorithm provides a probability which converges to the Boltzmann distribution. 

Among the many fields of application, the Metropolis algorithm has been widely used to investigate the equilibrium properties of polymeric chains, in particular of protein models. Important results were obtained by simulations of lattice model proteins concerning their free energy landscape \cite{dill,shakh}. In the following we shall focus our attention on chain models to describe proteins, although our results can be applied to other fields. 

Equation (\ref{masterequation}) describes a tailor--made dynamics which, in principle, has nothing to do with the actual dynamics of a polymer. The actual dynamics of the polymer is described (if one wants to describe implicitly the solvent) by Langevin's Equation \cite{vankampen}
\begin{equation}
\frac{dp}{dt}=F-\frac{\gamma}{m}p+\eta,
\end{equation}
where $p$ is the momentum of a given particle, $F$ is the force acting on it, $\gamma$ is the friction coefficient, $m$ the mass and $\eta$ a stochastic variable describing the interaction with the solvent. This variable satisfies $<\eta(t)>=0$ and $<\eta(t)\eta(t')>=D\gamma^2\delta(t-t')$, where the brackets indicate the average over the realizations of $\eta$ and $D$ is the diffusion coefficient. On the other hand, in implementing the Metropolis algorithm, one is free to choose any kind of fancy move, with the only goal of speeding up the sampling of conformational space. If the chosen move allows enormous jumps across the conformational space, it is clear that the resulting dynamics has nothing to do with the actual dynamics of the polymer. However, the set of moves chosen for protein models is usually quite realistic, involving mainly local moves of the atoms (e.g., flips, crankschafts, etc. \cite{anders}). Thus, one can ask whether the trajectories obtained with the Metropolis algorithm and small moves have some degree of realism, in the sense that they provide an approximation to the solution of Langevin's equation.

Both Langevin's equation and the Metropolis algorithm are stochastic, containing some randomness. Consequently, what makes physical sense is not a single trajectory but trajectories averaged over the randomness. Thus, in the following we will compare solutions of the Langevin's equation averaged over the force $\eta$ exerted by the solvent with the trajectories generated by the Metropolis algorithm, averaged over independent runs.

The relation between number of Monte Carlo steps and time was investigated in the case of spinoidal decomposition in two dimensions by Meakin and coworkers \cite{meakin} and for lattice gauge theory by Baillie and Johston \cite{baillie}.
For the simple case of a lattice model of a $\alpha$--helical hairpin, Rey and Skolnick showed \cite{rey} that the Metropolis simulations are independent on the set of moves chosen in the Metropolis algorithm and are consistent with Langevin dynamics simulations. In the present work, we want to investigate further on the conditions under which the average solution of Langevin's equation is well approximated by Metropolis algorithm for protein--like chains.

\section{The theory}

Following ref. \cite{vankampen}, one can show that, under the assumption that $p_{ap}$ allows only small transitions $\delta x\equiv x'-x$, the master equation (\ref{masterequation}) solved by the Metropolis algorithm approximates a diffusive Fokker--Plank equation, which is equivalent to Langevin's equation. In fact, writing $w(x,\delta x)\equiv w(x\rightarrow x')$, one can approximate
\begin{eqnarray}
w(x-\delta x,\delta x)p(x-\delta x,t)& \approx & w(x,\delta x)p(x,t)-\delta x\frac{\partial}{\partial x}\left(w(x,\delta x)p(x,t)\right)+ \nonumber\\
  &+& \frac{1}{2}(\delta x)^2\frac{\partial^2}{\partial x^2}\left(w(x,\delta x)p(x,t)\right),
\end{eqnarray}
so that the master equation (\ref{masterequation}) becomes
\begin{eqnarray}
\frac{\partial p(x,t)}{\partial t}&\approx&\int w(x,\delta x)p(x,t)d(\delta x) - \int \delta x\frac{\partial}{\partial x}[w(x,\delta x)p(x,t)]d(\delta x)+\nonumber\\
&+&\frac{1}{2}\int (\delta x)^2\frac{\partial^2}{\partial x^2}[w(x,\delta x)p(x,t)]d(\delta x)-\int w(x,\delta x)p(x,t)d(\delta x).\nonumber
\end{eqnarray}
and, since the first and last terms cancel each other, we get
\begin{equation}
\frac{\partial p(x,t)}{\partial t}=-\frac{\partial}{\partial x}\left(A(x)p(x,t)\right)+\frac{1}{2}\frac{\partial^2}{\partial x^2}\left(B(x)p(x,t)\right),
\end{equation}
where $A(x)\equiv \int d(\delta x)\;\delta x\,w(x,\delta x)$ and $B(x)\equiv \int d(\delta x)\;\delta x^2\,w(x,\delta x)$. Note that $A(x)$ and $B(x)$ are nothing else but the average displacement and the average square displacement of the coordinate $x$ (in other words, $<\delta x>$ and $<\delta x^2>$, respectively).

It is then enough to show that, among all possible Fokker--Plank equations, the one which rules the Monte Carlo sampling is that associated with diffusion in a potential $U$. This is true if $A(x)=-U'(x)/\gamma$ and $B(x)=2D$. Using the jumping rate of the Metropolis algorithm (\ref{metropolis}) it is possible to calculate the values of $A(x)$ and $B(x)$, using the scheme developed in ref. \cite{giapp}. To achieve this result, use is made of the hypothesis that $p_{ap}(x,\delta x)\equiv p_{ap}(x\rightarrow x')$ allows only small jumps. We define the small number $R$ as the maximum displacement allowed to the coordinate $x$ in a single move, and we assume that $R$ is independent on $x$ (in order to obtain, at the end, a homogeneous diffusion coefficient). For sake of simplicity, we can chose $p_{ap}(\delta x)$ equal to $(2R)^{-1}$ if $-R<\delta x<R$ and zero otherwise. Moreover, again due to the small jumps hypothesis, we can expand the jumping rate in
\begin{equation}
w(x,\delta x)=w_0 p_{ap}(\delta x)\min\left[ 1,\exp\left(-\frac{U(x+\delta x)-U(x)}{T}\right)\right]\approx w_0 p_{ap}(\delta x)\min
\left[ 1,1-\frac{U'(x)\delta x}{T}\right]
\end{equation}
In general, at $\delta x=0$ the sign of $U(x+\delta x)-U(x)$ changes, and the minimum function switches from the value $1$ to the exponential. We can thus break the integral of the definition of $A(x)$ at $\delta x=0$
\begin{eqnarray}
A(x)&=&\int_{-\infty}^\infty \delta x w(x,\delta x) d(\delta x)=\nonumber\\
&=& \frac{w_0}{2R}\int_{-R}^{0}\delta x d(\delta x)+\frac{w_0}{2R}\int_{0}^{R}\delta x d(\delta x)- \frac{w_0}{2R}\int_{0}^{R}\delta x^2\frac{U'(x)}{T} d(\delta x).\nonumber
\end{eqnarray}
The first two integrals cancel each other because they can be merged into the integral of an odd function on an even interval. Consequently,
\begin{equation}
A(x)=-w_0\frac{U'(x)\cdot R^2}{6T}.
\label{eq_a}
\end{equation}
In the same way one can calculate
\begin{eqnarray}
B(x)&=&\int_{-\infty}^\infty \delta x^2 w(x,\delta x) d(\delta x)=\nonumber\\
&=& \frac{w_0}{2R}\int_{-R}^{0}\delta x^2 d(\delta x)+\frac{w_0}{2R}\int_{0}^{R}\delta x^2 d(\delta x)- \frac{w_0}{2R}\int_{0}^{R}\delta^3 x\frac{U'(x)}{T} d(\delta x)=\nonumber\\
&=&w_0\frac{R^2}{3}-w_0\frac{|U'(x)|\cdot R^3}{8T}.
\end{eqnarray}
Since $R$ is small, the second term of the latter expression is negligible with respect to the former, so that
\begin{equation}
B(x)=w_0\frac{R^2}{3}
\label{eq_b}
\end{equation}
which is a constant. This corresponds to Langevin dynamics with $\gamma=6T/w_0R^2$ and $D=w_0R^2/6$ (these expressions will be commented in Sect. \ref{sect_temp}).

\section{Comparing Langevin and Metropolis simulations}

The above derivation show that the Metropolis algorithm can be used to solve Langevin's equation, provided that the allowed moves are small. On the other hand, it gives no indication about how small the move must be. To investigate this point, we have compared the average trajectories generated by the Metropolis algorithm with numerical solutions of Langevin's equation. 

The protein model used in the following is a simplified G\=o model \cite{go}. G\=o models, at different degrees of geometric resolution, have been widely used in the literature to investigate thermodynamical and kinetic properties of proteins. In their C$_\alpha$-version they allow to perform massive simulations with small-- to medium--sized proteins \cite{go1,go2,go3,go4}. Another choice is to account also for the side--chain, as a single bead \cite{go5,go6}, in order to give a more realistic description of the protein without increasing much the computational cost. In fact, with this model it was possible not only to simulate the folding, but also the aggregation of a number of identical proteins \cite{go7}. Another possible choice is to give a full description of the atomic structure of amino acids, where the basic interacting unit is the atom \cite{go8,go9} (for a careful review see ref. \cite{ding}). In order to perform the simulations we have employed a C$_\alpha$ G\=o--model, where each amino acid is described as a spherical bead.  Two amino acids which are in contact (whose $C_\alpha$ distance is $< 6.5 \rAA$ in the experimental native conformation and which are not consecutive along the chain) interact through a 6--12 Lennard--Jones potential, whose bottom lies at $R_0=0.8\times d_{ij}^N$, and energy $-2$ kcal/mol and whose cutoff is at 20$\rAA$. The other pairs of amino acids repell each other with a $(4.5/r)^{12}$ potential.

In the implementation of the Metropolis algorithm, at each step an amino acid is chosen at random with flat probability. The chosen amino acid is rotated of an angle $\Delta\alpha$ around the axis defined by the previous and the following one, where $\Delta\alpha$ is a random number generated from a flat distribution and constrained in the interval $-\alpha_M<\Delta\alpha<\alpha_M$. This set of moves varies the dihedrals and the angles associated with the amino acids, while the bonds between consecutive amino acids are inextensible and display length $R_{b}=3.84\rAA$. The first and last amino acid of the chain diplay a further degree of freedom, that is the bond angle with the previous amino acid, which can be also moved at random.

The Langevin equations of motion are integrated with the B.B.K. algorithm \cite{bbk} using a time step $\Delta t=10fs$. Each amino acid is linked to the consecutive ones through a harmonic potential with spring constant $k=10 Kg/s^2$. This is the only difference between the model used for the molecular dynamics and that used for the Metropolis simulations. Due to the stiffness of the spring, we do not expect that this produces relevant effects. 

\begin{figure}
\centerline{\includegraphics*[width=12cm]{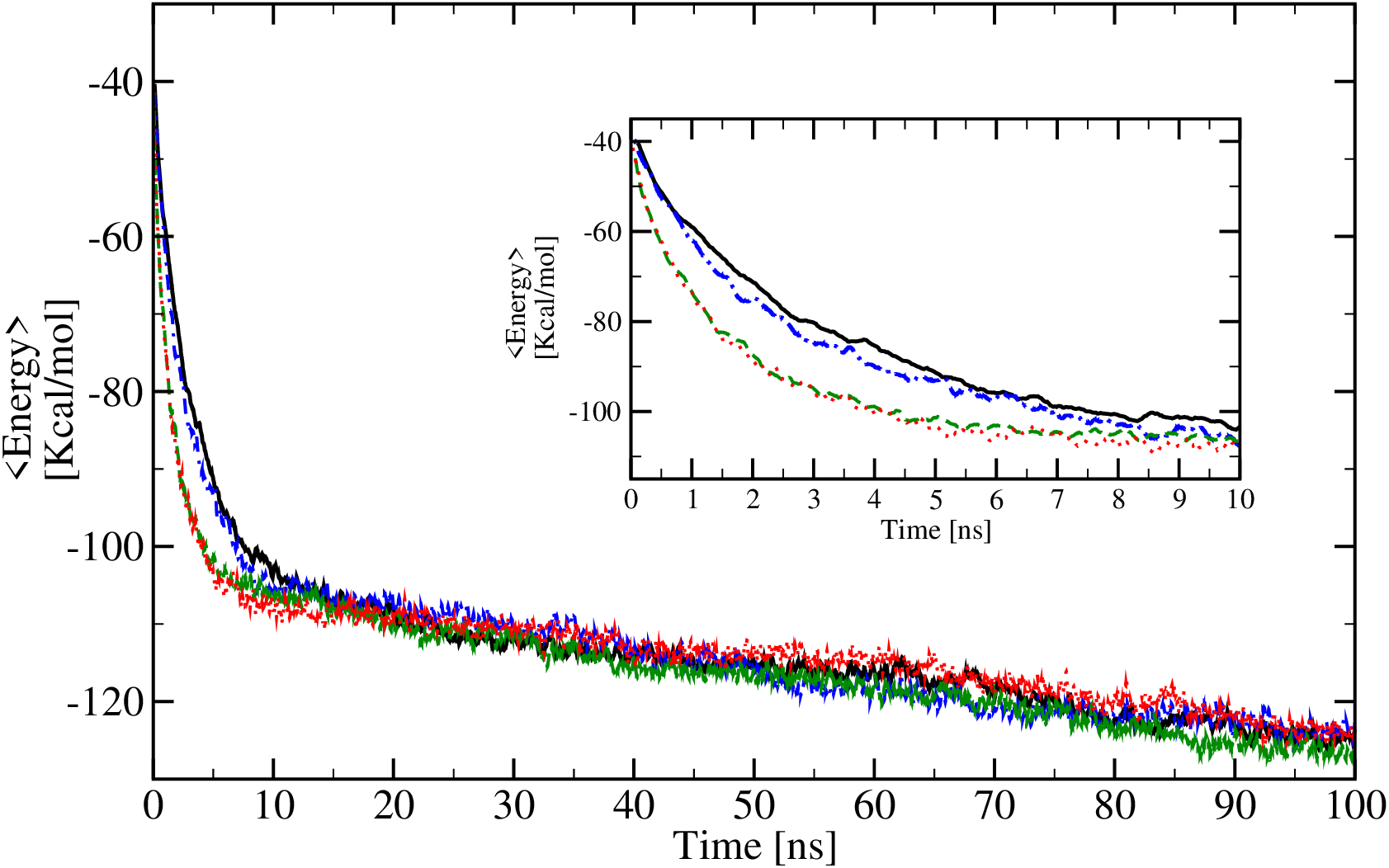}}
\caption{The mean energy as a function of time for the Langevin simulation (continuous curve) and for three Monte Carlo simulations with different values of the maximum allowed dihedral angle move: $\alpha_M=0.5$ (dot dashed blue curve), $\alpha_M=4$ (dashed green curve), $\alpha_M=180$ (dotted red curve). The inset shows a zoom in of the first 10 ns.}
\label{fig1}
\end{figure}

In order to compare Langevin and Metropolis dynamics, we have performed a number of simulations at temperature $T=400 \deg K$, lower than the folding temperature ($T_f=440 \deg K$), with both methods on a short protein (60 residues), the SH3 domain of Src, starting from random generated conformations. In order to evaluate the dependence of the results on the length of the protein, we will also study betanova, a synthetic peptide built out of 18 residues and the monomer of HIV--PR, a protein of 99 residues. The solid curve in Fig. \ref{fig1} is the energy of SH3 as a function of time calculated with the Langevin dynamics, averaged over 50 independent runs. Metropolis simulations have been performed at the same temperature, using different values of the threshold $\alpha_M$ of the angular move. For each of them, we have calculated the average energy as a function of the number of Monte Carlo steps. To transform the number of Monte Carlo steps (MCs) into time, we have calculated the time content $\tau_0$ of a single Monte Carlo step ($\tau_0\equiv 1/w_0$), optimizing its value by means of a least--square minimization of the curve obtained from the Monte Carlo simulation versus that obtained from the Langevin dynamics. Three examples of the resulting curves are also displayed in Fig. \ref{fig1} with dotted ($\alpha_M=180\deg$), dashed ($\alpha_M=4\deg$) and dotted--dashed ($\alpha_M=0.5\deg$) lines. 

\begin{figure}
\centerline{\includegraphics*[width=12cm]{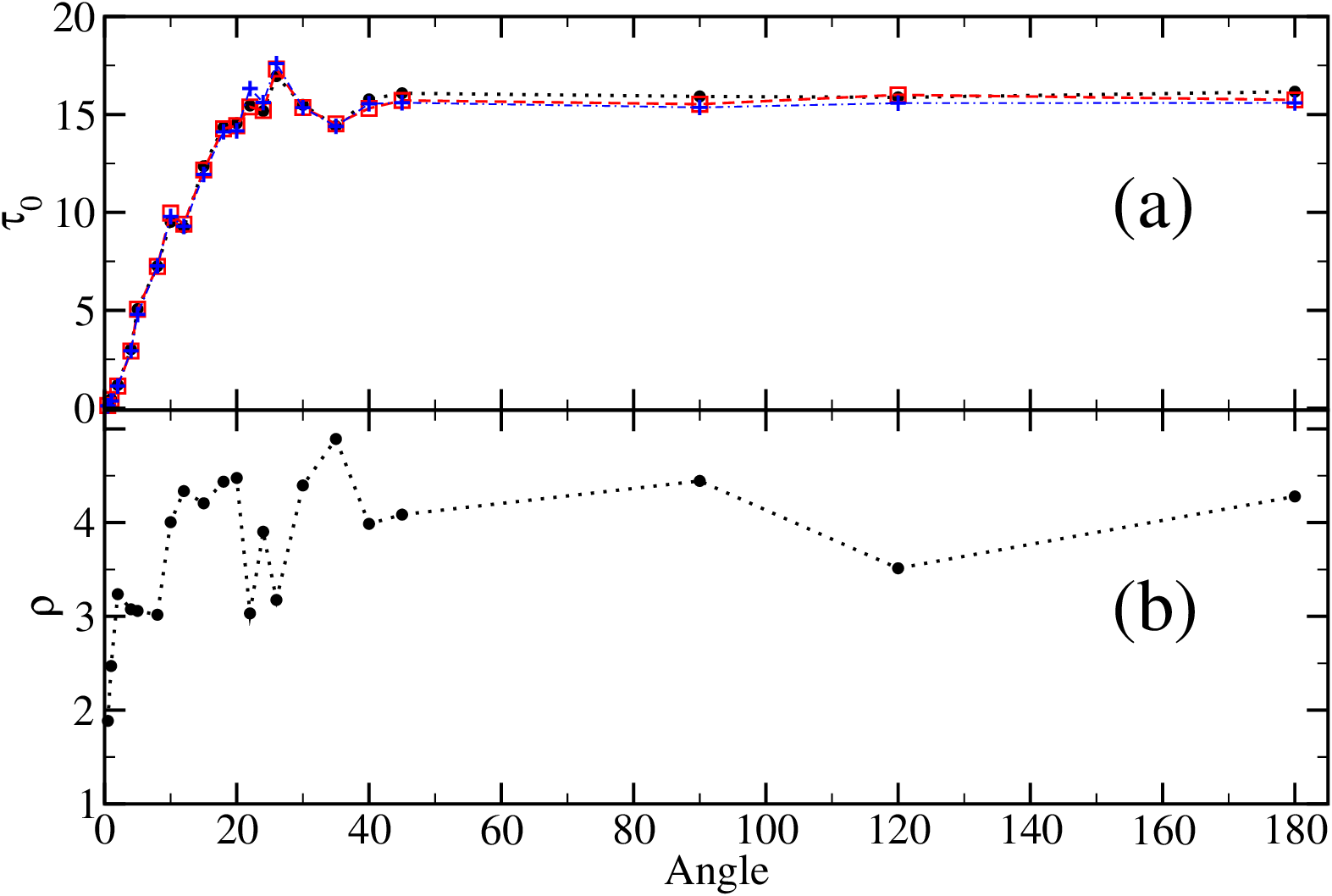}}
\caption{(a) The time content of a Monte Carlo step $\tau_0$ ($\equiv 1/w_0$) calculated by means of a least--square minimization of the curve of the energy (black dots), of the dRMSD (blue cross) and of the parameter q (red squares). (b) the value of the square--root of the average of residuals $\rho$ as a function of the maximum allowed dihedral angle move used in Monte Carlo simulation.}
\label{fig2}
\end{figure}

The values of $\tau_0$ and of the square--root of the average of residuals $\rho$ obtained from the optimization are displayed in Fig. \ref{fig2}. These plots show that in order to have in the out--of--equilibrium regime an error in the determination of the energy smaller than few kcal/mol (i.e., a typical single non--bonded interaction in biological molecules) one needs to set $\alpha_M\leq 1\deg$. Larger values of $\alpha_M$ lead to errors of the order of 4-5 kcal/mol in the whole out--of--equilibrium phase (cf. Fig. \ref{fig2}), which can reach 10 kcal/mol in the first nanoseconds (cf. Fig. \ref{fig1}). Of course, in the long--time limit the curves overlap, by virtue of the ergodic theorem.

\begin{figure}
\centerline{\includegraphics*[width=12cm]{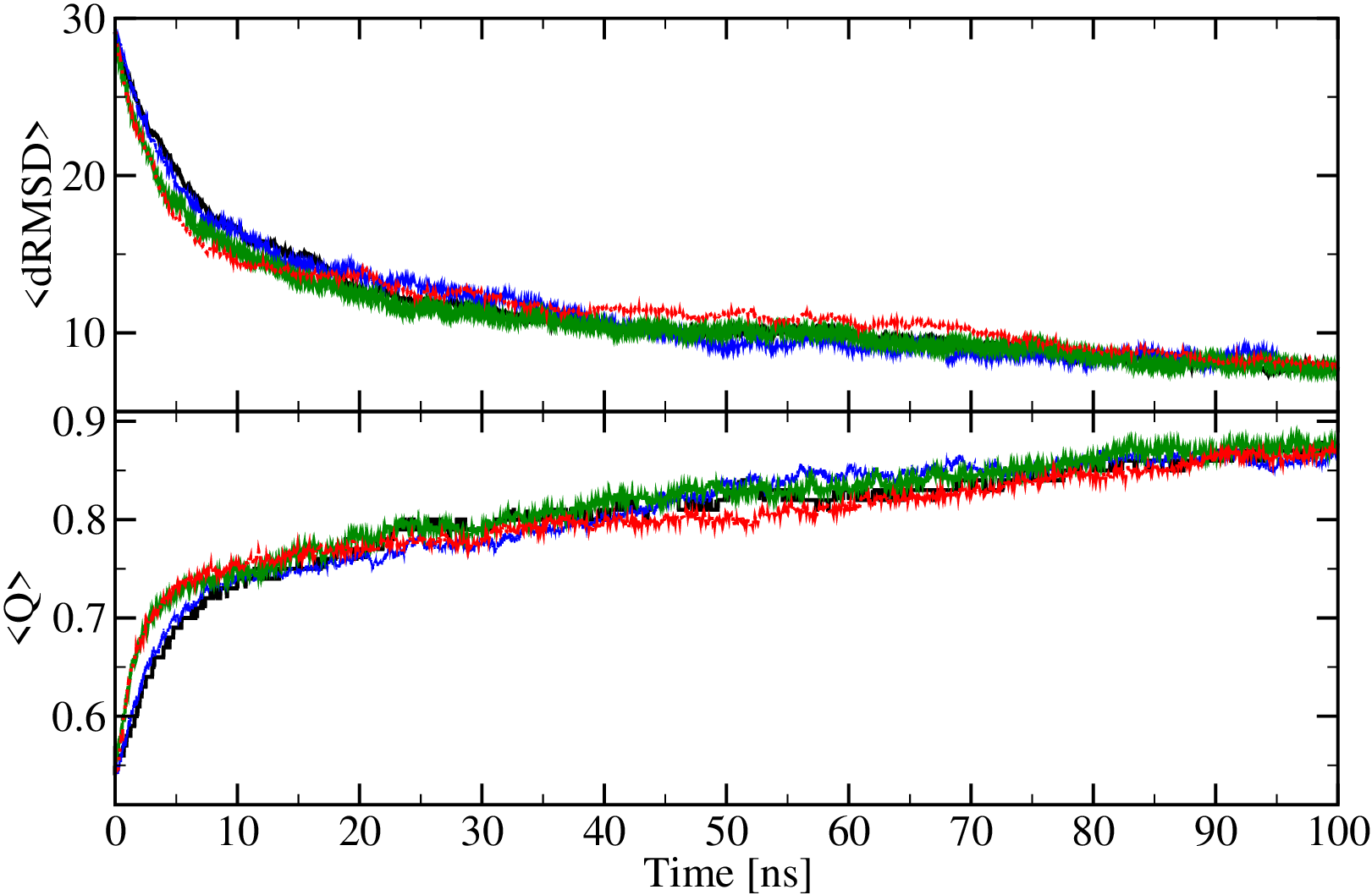}}
\caption{Same as Fig. \protect\ref{fig1} but for the mean value of the dRMSD and of the parameter q.}
\label{fig3}
\end{figure}

In Fig. \ref{fig3} is displayed the time dependence of the mean dRMSD and $q$, two order parameters which indicate to which extent a conformation of the protein chain is similar to the native conformation. The dRMSD is defined as $[N^{-2}\sum_{i<j}(d_{ij}-d_{ij}^N)^2]^{1/2}$, where $N$ is the number of amino acids of the chain, $d_{ij}$ is the distance between two of them in the current conformation and $d_{ij}^N$ is the same quantity calculated in the native conformation. The parameter $q$ is the fraction of contacts that a given conformation shares with the native conformation, having defined two amino acids to be in contact if they are closer than 6.5$\rAA$. These curves show the same behaviour of the energy, a good description being only provided for $\alpha_M\leq 1\deg$.

The time content of a Monte Carlo step in the most reliable case $\alpha_M=0.5\deg$ is 1MCS = 0.1 fs, and it increases almost linearly up to 1MCS = 15 fs at $\alpha_M=30\deg$, where it reaches a plateau (see Fig. \ref{fig2}). A time step of 0.1 fs is quite small, smaller than that usually employed in molecular dynamics simulations. However, this limit is compensated by the fact that Metropolis algorithm are often computationally much faster than molecular dynamics algorithm. Moreover, if one requires a less stringent description of the initial stages of the dynamics, it is possible to use a larger threshold $\alpha_M$. To be noted that the above relationship holds for a potential shaped with a Lennard--Jones function of the kind used in these calculations, and thus will be more favourable for smoother (although, possibly, less realistic) potentials.

\begin{figure}
\centerline{\includegraphics*[width=12cm]{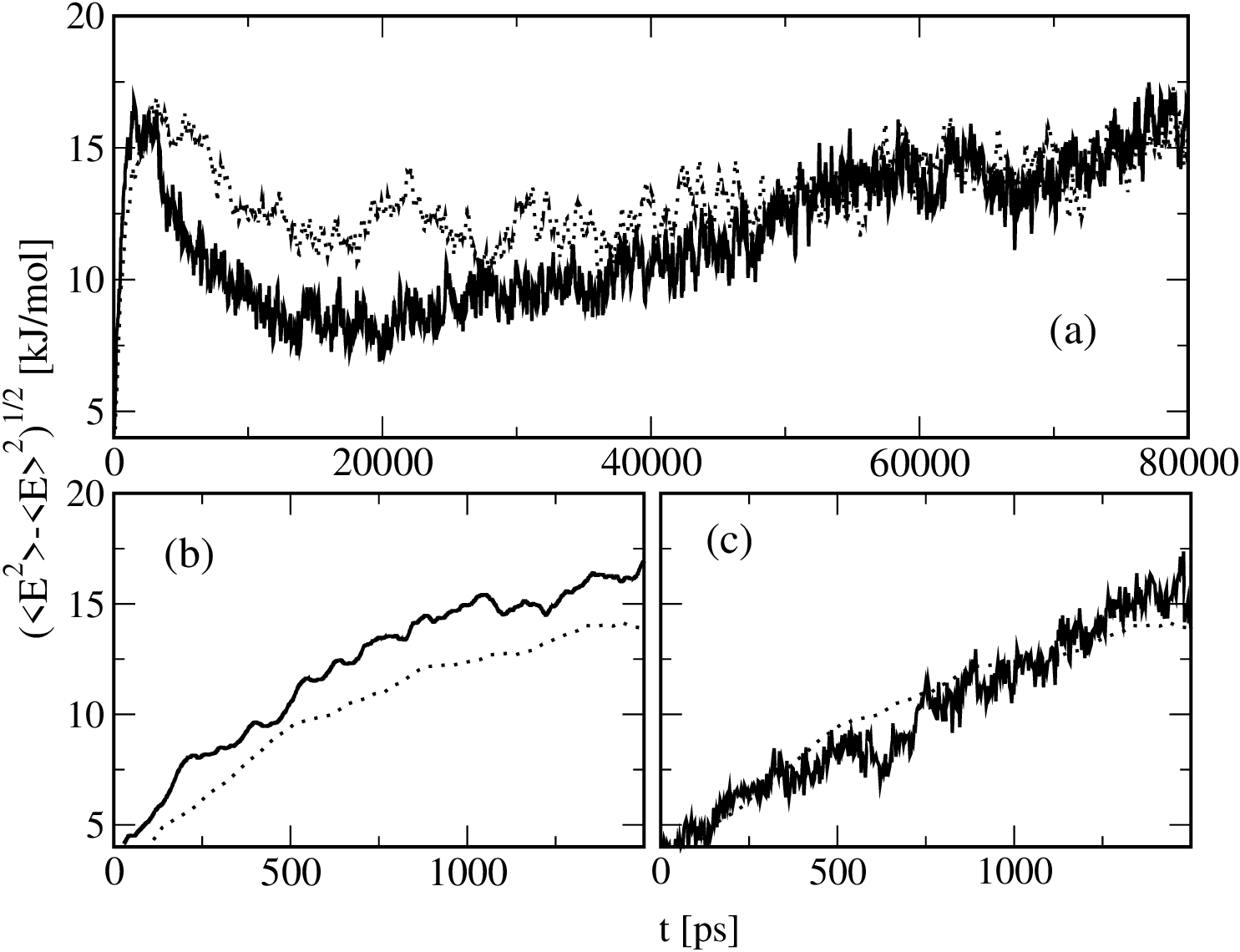}}
\caption{(a) The energy fluctuations obtained by the Monte Carlo with $\alpha_M=0.5$ (continuous curve) and Langevin simulations (dotted curve). (b) A zoom of (a). (c) The energy fluctuations obtained setting $\alpha_M=0.05$.}
\label{fluct}
\end{figure}

We have also compared the energy fluctuations $(<E^2>-<E>^2)^{1/2}$ obtained by the Monte Carlo and Langevin simulations. The results obtained for $\alpha_M\leq 0.5\deg$ are reported in Fig. \ref{fluct}(a). Although the overall behaviour is quite similar, the Monte Carlo simulations understimates the fluctuations in the first nanoseconds of up to 3 kcal/mol (i.e., $\approx 5 kT$). As the system approaches equilibrium (cf. the last tens of nanoseconds), the two curves overlap better, as expected by virtue of the ergodic theorem. In order to obtain a better overlap, it is necessary to further reduce the value of $\alpha_M$. In Fig. \ref{fluct}(c) are displayed the same quantities as above, but obtained using $\alpha_M\leq 0.05\deg$. Note that we could simulate only the first 1.5 ns, due to the much larger computational cost of choosing such a small elementary move. However, in this case the match between the fluctuations obtained with the Monte Carlo algorithm and the Langevin dynamics is much better (root mean square residual is 1.1 kcal/mol) than in the case where we used $\alpha_M\leq 0.5\deg$ on the same time interval (root mean square residual is 3.1 kcal/mol, cf. Fig. \ref{fluct}(b)).

Summing up, one can conclude that the Metropolis dynamics is a fast algorithm to describe the dynamics of mean quantities, but is not useful if one requires high precision ($<kT$) in the higher moments. 

\section{Dependence on the temperature} \label{sect_temp}

The link between Metropolis and Langevin dynamics, provided by Eqs. \ref{eq_a} and \ref{eq_b}, displays the unrealistic feature that the effective friction coefficient $\gamma$ results linearly dependent on the temperature of the system ($\gamma=6T/w_0R^2$), while the diffusion coefficient $D$ results independent on $T$ ($D=w_0R^2/6$). Nonetheless, these two quantities satisfy Einstein's relation $D=T/\gamma$. The independence of $\gamma$ and the linear dependence of $D$ on the temperature can be reached if one assumes that the value of $w_0$ associated with the Metropolis step varies linearly with the temperature, i.e. $w_0(T)=w'_0\cdot T$, where the constant $w'_0$ is independent on $T$. This provides the correct $\gamma=6/w'_0R^2$ and $D=Tw'_0R^2/6$.

\begin{figure}
\centerline{\includegraphics*[width=12cm]{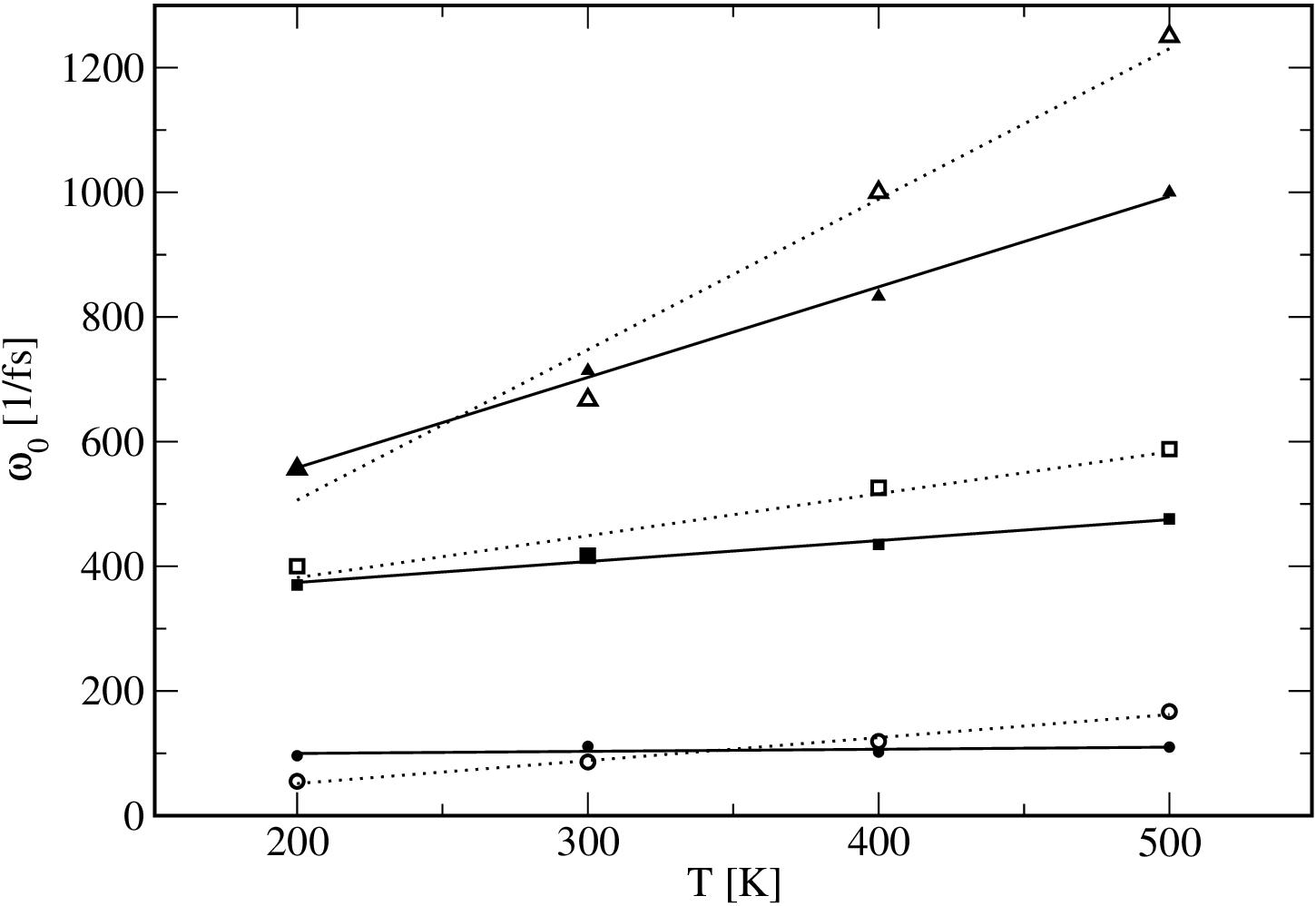}}
\caption{The values of $w_0$ obtained from the fit of the energy as a function of temperature for HIV--PR (filled triangles), SH3 (filled squares) and betanova (filled circles). The open symbols indicate the values of $w_0$ obtained by the fit of the displacement of the centre of mass. The straight lines indicate the linear fit.}
\label{fig_T}
\end{figure}

To check numerically this result, we have performed both Metropolis (using $\alpha_M=0.05\deg$) and Langevin simulations at various temperatures ranging from $T=200$K to $T=500$K. The values of $w_0$ obtained from the simulations of SH3 (following the same procedure described in the previous Section) are displayed with filled squares in Fig. \ref{fig_T} as a function of temperature. The correlation coefficient of the linear fit $w_0(T)=w'_0\cdot T$ is $0.987$, indicating a good linear behaviour (see also Table \ref{table_T}). Note that the low value of $w'_0$ suggests that for proteins of the size of SH3 and for temperature variations within the range of biological relevance, the error done assuming a $w_0$ independent of the temperature is small.

In addition, we have also studied the displacement of the centre of mass of the protein, calculating the diffusion coefficient both in the case of Langevin and Metropolis simulations. The value of $w_0$ obtained from the comparison of the diffusion coefficients is also displayed in Fig. \ref{fig_T} (empty squares) and the linear coefficient $w''_0$ of the fitting line is listed in Table \ref{table_T}. One would have expected $w''_0$ to be identical to $w'_0$. Their difference suggests that the small approximation done in each Metropolis step with respect to the Langevin dynamics sums up to displace the centre of mass of the protein, giving rise to a less precise approximation. 

\section{Dependence on the length of the protein}

The results discussed above are obtained with a protein composed of 60 amino acids. In order to elucidate to which extent these results depend on the length $N$ of the protein, we have repeated the same kind of molecular dynamics and Monte Carlo simulations for other two proteins, that is betanova ($N=18$) and the monomer of HIV--1 protease ($N=99$).

\begin{figure}
\centerline{\includegraphics*[width=12cm]{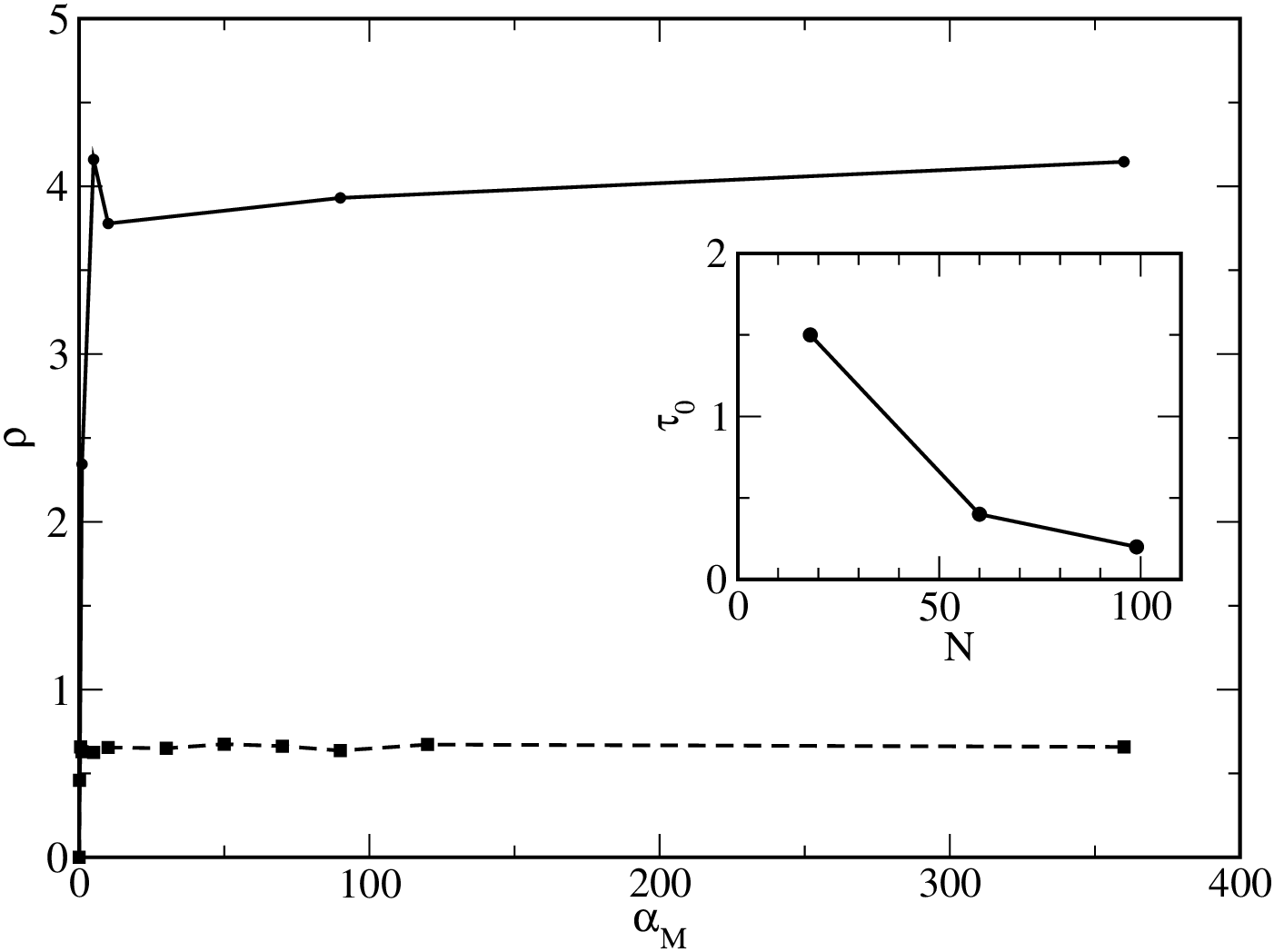}}
\caption{The square--root of the average of residuals $\rho$ as a function of the maximum allowed dihedral angle move used in Monte Carlo simulation for the HIV--1 protease (solid curve) and betanova (dashed curve). In the inset, the value of $\tau_0$ at $\alpha_M=1$ for the three proteins, as a function of their length $N$.}
\label{fig_N}
\end{figure}

The mean square energy deviation between the two kind of simulations as a function of $\alpha_M$ are plotted in Fig. \ref{fig_N} with a solid curve for HIV--1 Protease and with a dashed curve for betanova. While HIV--1 protease behaves similarly to SH3 (cf. Fig. \ref{fig2}), betanova displays low deviations even for large angles $\alpha_M$. In any case, at low values of $\alpha_M$, all of them display small values of $\rho$. For example, at $\alpha_M=1\deg$ the values of $\rho$ are $2.34$ for HIV--1 protease (i.e., $N=99$), $2.47$ for SH3 ($N=60$) and 0.62 for betanova ($N=18$). Consequently, a threshold of $\alpha_M=1\deg$ is a safe choice for most cases of interest.

The variation of $\tau_0$ ($\equiv 1/w_0$) with respect to $N$ is displayed in the inset to Fig. \ref{fig_N}. Again, the shortest protein displays a behaviour which is quite different from the other two, having a $\tau_0$ which is an order of magnitude larger. On the other hand, the two larger proteins display values of $\tau_0$ of $0.2$ and $0.4$fs, respectively, that is of the same order of magnitude.

The dependence of $w_0$ on the temperature for the different protein sizes is displayed in Fig. \ref{fig_T} and the coefficients $w''_0$ of the linear fit summarizd in Table \ref{table_T}. These results show that $w''_0$ increases quite rapidly as the length of the protein increases. Thus one can conclude that in Metropolis simulations of proteins larger than SH3 it becomes necessary to account explicitely for the dependence of $w_0$ on the temperature.

\section{Conclusions}

We have shown that Metropolis algorithm can be used to simulate the dynamics of a simplified protein model, provided that the residue are moved of small dihedrals and that the probability of chosing a move is independent of the conformation of the chain. If one is investigating mean quantities, a constrain of $1\deg$ on the dihedral move is enough. This approach can be useful not only because it gives a physical meaning to the trajectories obtained by Monte Carlo thermodynamical samplings, but also because for simple protein models the Metropolis algorithm can be faster than Langevin's dynamics. For example, a 1--ns simulation of the model chain at $T=400$ and with $\alpha_M=1\deg$ took, with our codes and on the same PC, 8.5 s when performed with Metropolis algorithm and 16.0 s when performed with Langevin's algorithm. Of course, the possibility to use Metropolis simulations to study the dynamics of a protein implies the knowledge of the conversion factor $\tau_0$. If one is interested in the precise value of $\tau_0$, some molecular dynamics runs are necessary in order to estimate it, and consequently the computational advantage of using Metropolis simulations is lost. If, on the other hand, one needs only the order of magnitudes of the folding time, one can use the conversion factors obtained above (i.e., $\tau_0$ of the order of $0.5$ fs at room temperature).

\acknowledgments{G. T. acknowledges the financial support of the 2003 FIRB program of the Italian Ministry for Univeristy and Scientific Research.}

\newpage

\begin{table}
\begin{tabular}{|l|l|l|l|l|l|}\hline
protein & length & $w'_0$ & $r$ & $w''_0$ & $r$ \\\hline
betanova & 18 & 0.033 & 0.601 & 0.369 & 0.994 \\\hline
SH3 & 60 & 0.336 & 0.987 & 0.673 & 0.968 \\\hline
HIV--PR & 99 & 1.451 & 0.998 & 2.451 & 0.984 \\\hline
\end{tabular}
\caption{The linear coefficient $w'_0$ (in $(fs \cdot K)^{-1}$) which controls the dependence of $w_0$ on the temperature for the three proteins, characterized by different lengths (third column) and the associated correlation coefficient $r$ (fourth column). The value of $w''_0$ obtained comparing the displacement of the centre of mass is also listed in the fifth and sixth column.}
\label{table_T}
\end{table}

\newpage

\end{document}